\documentclass[12pt,a4paper]{article}
\usepackage[utf8]{inputenc}
\usepackage{latexsym}
\usepackage{amsfonts,amsmath,amssymb}
\usepackage{fancyhdr}
\usepackage{proba}
\usepackage{graphics,graphicx}
\usepackage{setspace}
\usepackage{float}
\usepackage{multirow}
\usepackage{pifont}
\usepackage{authblk}
\usepackage{multirow}
\usepackage{natbib}
\usepackage{microtype}
\usepackage{textcomp}
\usepackage{xcolor}
\newcommand*{\bff}[1]{\ifmmode\pmb{#1}\else\textbf{#1}\fi}

\usepackage{hyperref}

\title{Bayesian inference for the Net Promoter Score}
\author[1]{Eliardo G. Costa\thanks{eliardo.costa@ufrn.br}}
\author[2]{Rachel Tarini Q. Ponte}
\affil[1]{Departamento de Estatística, Universidade Federal do Rio 
Grande do Norte, Natal/RN, Brazil}
\affil[2]{Ânima Educação, Natal/RN, Brazil}

\begin{document}

\maketitle
\onehalfspacing

\begin{abstract}
The Net Promoter Score is a simple measure used by several companies as indicator of customer loyalty. Studies that address the statistical properties of this measure are still scarce and none of them considered the sample size determination problem. We adopt a Bayesian approach to provide point and interval estimators for the Net Promoter Score and discuss the determination of the sample size. Computational tools were implemented to use this methodology in practice. An illustrative example with data from financial services is also presented.
\\

\end{abstract}

\section{Introduction}

\cite{Reichheld2003} proposed a statistics called Net Promoter Score (NPS) that may be used by a company as an indicator of customer loyalty. The author applied a questionnaire with some questions related to loyalty to a sample of customers of some industries, and with the purchase history of each customer it was possible to determine which questions had the strongest statistical correlation with repeat purchase or referrals. One of these questions performed better in most industries: ``How likely is it that you would recommend [company X] to a friend or colleague?''. \cite{Reichheld2003} suggested that the response to the this questions must be on a 0 to 10 rating scale. Then, it is considered ``promoters'' the customers who respond with 9 or 10, ``passives'' the customers who respond with 7 or 8, and ``detractors'' the customers who respond with 0 through 6. The idea is that the more ``promoters'' company X has, the bigger its growth. An estimate of the NPS is computed as the difference between the proportions (or percentages) of ``promoters'' and ``detractors''. \cite{Keininghametal2008} discuss the claims that NPS is the single most reliable indicator of a company's ability to grow, and that it is a superior metric to costumer satisfaction. \cite{Rocks2016} presents a brief summary of some critiques about the NPS, see references therein.



In the context of statistical modeling, \cite{Rocks2016} focus on estimating intervals for the NPS in a frequentist approach via Wald intervals and Score methods. Also, the author perform a study simulation to assess the coverage probability of the proposed interval estimates, and conclude that variations on the adjusted Wald and an iterative Score method performed better. \cite{Markoulidakisetal2021} approach the customer experience as a NPS classification problem via machine learning algorithms. We may also cite \cite{EskildsenKristensen11} and \cite{KristensenEskildsen14} for related work. Studies that address the statistical properties of this measure are still scarce and none of them, to the best of our knowledge, considered the sample size determination problem. In this context, we propose a Bayesian model in order to make inference for the NPS and to establish a sample size determination methodology. See \cite{RossiAllenby03} for an exposition of the usefulness of the Bayesian methods in marketing.


In Section \ref{sec-bayes-model}, we describe the Bayesian model and the methodologies to obtain point and interval estimates for the NPS. The problem of the minimum sample size determination is discussed and implemented in Section \ref{sec-min-n}. In Section \ref{sec-example} we present an illustrative example with data on financial services. We conclude with some remarks in Section \ref{sec-remarks}.2

\section{Bayesian model}\label{sec-bayes-model}

Let $\bff{\theta}=(\theta_1,\theta_2,\theta_3)$, where $\theta_1, \theta_2$ and $\theta_3$ are the proportions of detractors, passives and promoters in the customer population, respectively. Then, the NPS in the respective population is given by $\Delta=\theta_3-\theta_1$, the parameter of interest. In a sample of $n$ customers we count the number of customers in each category based on their responses for the aforementioned question. Let $\bff{X}_n=(X_1,X_2,X_3)$, where $X_1, X_2$ and $X_3$ are numbers of customers categorized as detractors, passives and promoters, respectively, in the customer sample.

Given $\bff{\theta}$, we assume a multinomial distribution for the counts $\bff{X}_n$, and we denote $\bff{X}_n|\bff{\theta}\sim\text{Mult}(n, \bff{\theta})$. The respective probability distribution is given by
$$\probX{X_1=x_1,X_2=x_2,X_3=x_3}=\frac{n!}{x_1!x_2!x_3!}\theta_1^{x_1}\theta_2^{x_2}\theta_3^{x_3},$$

\noindent
where $x_1,x_2,x_3=0,1,\ldots, n$ such that $x_1+x_2+x_3=n$, and $\theta_1+\theta_2+\theta_3=1$.

The natural (conjugate) choice for the prior distribution of $\bff{\theta}$ is a Dirichlet distribution, we denote $\bff{\theta}\sim\text{Dir}(\bff{\alpha})$ and the respective probability density function is given by
$$\pi(\bff{\theta})=\frac{\Gamma(\alpha_1+\alpha_2+\alpha_3)}{\Gamma(\alpha_1)\Gamma(\alpha_2)\Gamma(\alpha_3)}\theta_1^{\alpha_1-1}\theta_2^{\alpha_2-1}\theta_3^{\alpha_3-1},$$

\noindent
where $\theta_1+\theta_2+\theta_3=1$, $\bff{\alpha}=(\alpha_1, \alpha_2, \alpha_3)$ is a vector of positive hyperparameters and $\Gamma(\cdot)$ is the gamma function. The model may be written hierarchically as follows
\begin{equation}
 \bff{X}_n|\bff{\theta}\sim \text{Mult}(\bff{\theta});\quad
 \bff{\theta}\sim \text{Dir}(\bff{\alpha}).
\end{equation}

In this setting, given a observation $\bff{x}_n$ of $\bff{X}_n$, we have that the posterior distribution for $\bff{\theta}$ is a Dirichlet distribution with parameter $\bff{\alpha}+\bff{x}_n$, {\it i.e.}, $\bff{\theta}|\bff{x}_n\sim\text{Dir}(\bff{\alpha}+\bff{x}_n)$ \citep{Turkmanetal2019}. Also, Bayesian updating becomes straightforward since the current parameters of the posterior distribution may be used as the hyperparameters of the prior distribution in the next sampling of $\bff{X}_n$. Given a way to generate random values from the Dirichlet distribution, this provide us a simple way to draw values from the posterior distribution of $\Delta$ in order to obtain, approximately, posterior summaries as the mean, median, variance, quantiles, etc, and make inferences about the NPS.

An algorithm to obtain a sample of size $N$ from the posterior distribution of the $\Delta$ is outlined as follows.

\begin{enumerate}
 \item Set the values of $\bff{\alpha}$, $\bff{x}_n$ and $N$ ({\it e.g.}, $N=1000$).
 \item Draw a value of $\bff{\theta}=(\theta_1,\theta_2,\theta_3)$ from the Dirichlet distribution with parameter $\bff{\alpha}+\bff{x_n}$.
 \item Compute $\Delta=\theta_3-\theta_1$ and keep this value.
 \item Repeat Steps 2-3 $N$ times.
\end{enumerate}

It is well known that the marginal distributions of a Dirichlet distribution are beta distributions. Let $\bff{\alpha}^*=\bff{\alpha}+\bff{x}_n=(\alpha_1^*, \alpha_2^*, \alpha_3^*)^\top$. Then, it follows that
$$\theta_1|\bff{x}_n\sim\text{Beta}(\alpha_1^*, \alpha_2^*+\alpha_3^*)\quad\text{and}\quad\theta_3|\bff{x}_n\sim\text{Beta}(\alpha_3^*, \alpha_1^*+\alpha_2^*),$$
which give us that the mean of the posterior distribution of the NPS is
\begin{eqnarray}\label{post-mean}
 \cEX{\Delta}{\bff{x}_ n}=\cEX{\theta_3-\theta_1}{\bff{x}_n}=\frac{\alpha_3^*-\alpha_1^*}{\alpha_0^*},
\end{eqnarray}
where $\alpha_0^*=\alpha_1^*+\alpha_2^*+\alpha_3^*$. This mean may be used as a point estimator for the NPS. The respective variance is given by
\begin{eqnarray}\label{post-var}
 \cVarX{\Delta}{\bff{x}_n}&=&\cVarX{\theta_3}{\bff{x}_n}+\cVarX{\theta_3}{\bff{x}_n}-2\text{Cov}(\theta_3,\theta_1|\bff{x}_n)\nonumber\\
 &=&\frac{\alpha_1^*\alpha_2^*+\alpha_2^*\alpha_3^*+4\alpha_1^*\alpha_3^*}{(\alpha_0^*)^2(\alpha_0^*+1)}.
\end{eqnarray}

A credible interval that we may construct is based on (\ref{post-mean}) and (\ref{post-var}), {\it i.e.}, $\cEX{\Delta}{\bff{x}_ n}\pm \gamma\sqrt{\cVarX{\Delta}{\bff{x}_n}}$, where $\gamma$ is a fixed constant. We developed an Excel spreadsheet that computes this credible interval and a point estimate based on (\ref{post-mean}) and (\ref{post-var}). See Supplementary Material for more details.

Another credible interval may be specified by the highest posterior density (HPD) interval. In this case we use a Monte Carlo approach to approximate the HPD interval. In other words, we use a sample drawn from the posterior distribution of $\Delta$, which may be easily done since the posterior distribution is a Dirichlet distribution. See \citet[][pgs. 47-48]{Turkmanetal2019} for more details.

\section{Minimum sample size}\label{sec-min-n}

To determine the minimum sample size required to estimate $\Delta$ with a pre-specified precision, we consider a criterion based on the average length of credible intervals. The posterior credible interval accounts for the magnitude of the NPS and this may help the company to know when to perform a gap analysis and create a business action plan in order to improve the NPS, {\it i.e.}, increase the NPS until the company has more promoters than detractors ($\Delta>0$).


Let $a(\bff{x}_n)$ and $b(\bff{x}_n)$ be the lower and upper bounds of the HPD interval for $\Delta$. The rationale here is to set the minimum Bayesian coverage probability $1-\rho$ and obtain the minimum sample size by requiring that the length of the HPD interval $\ell(\bff{x}_n)=b(\bff{x}_n)-a(\bff{x}_n)$ be such that
\begin{equation}
 \int_\mathcal{X}\ell(\bff{x}_n)g(\bff{x}_n)\,d\bff{x}_n\leq\ell_{\text{max}},
\end{equation}
where $\ell_{\text{max}}$ is the maximum admissible length for the HPD interval, $\mathcal{X}$ is the sample space associated to $\bff{x}_n$ and $g(\bff{x}_n)$ is the marginal probability function of the outcomes. This is called average length criterion (ALC). See \cite{Costaetal21} and references therein for more details about this criterion.

Since it is impractical to obtain analytically the lower and upper bounds of the HPD interval for $\Delta$, we use a Monte Carlo approach \citep{ChenShao99} to obtain the respective bounds as well as the the respective integral.


An algorithm to obtain the minimum sample size satisfying this criterion is outlined as follows.

\begin{enumerate}
 \item Set values for $\ell_{\text{max}}$, $\bff{\alpha}$, $\rho$ and take $n=1$.
 \item Draw a sample of size $L$ ({\it e.g.}, $L=1000$) of $\bff{x}_n$; to draw $\bff{x}_n$, first draw one value of $\bff{\theta}$ from the Dirichlet distribution with parameter $\bff{\alpha}$ and given this value, draw $\bff{x}_n$ from the multinomial distribution with parameter $\bff{\theta}$. 
 \item Obtain the HPD interval of probability $1-\rho$ for each $\bff{x}_n$ that was drawn and then the respective interval length: for each value drawn in Step 2, obtain the lower and upper bounds of the HPD interval of probability $1-\rho$ as indicated in \cite{ChenShao99}. Then, compute the difference between the upper and lower bounds for each value drawn in order to obtain the interval lengths.
 \item Compute the average of the $L$ HPD interval lengths.
 \item If this average is lower or equal to $\ell_{\text{max}}$, stop. The value $n$ obtained in this step is the required value. Otherwise, set $n=n+1$ and return to Step 2.
\end{enumerate}

\begin{figure}[h]
 \centering
 \includegraphics[scale=0.32]{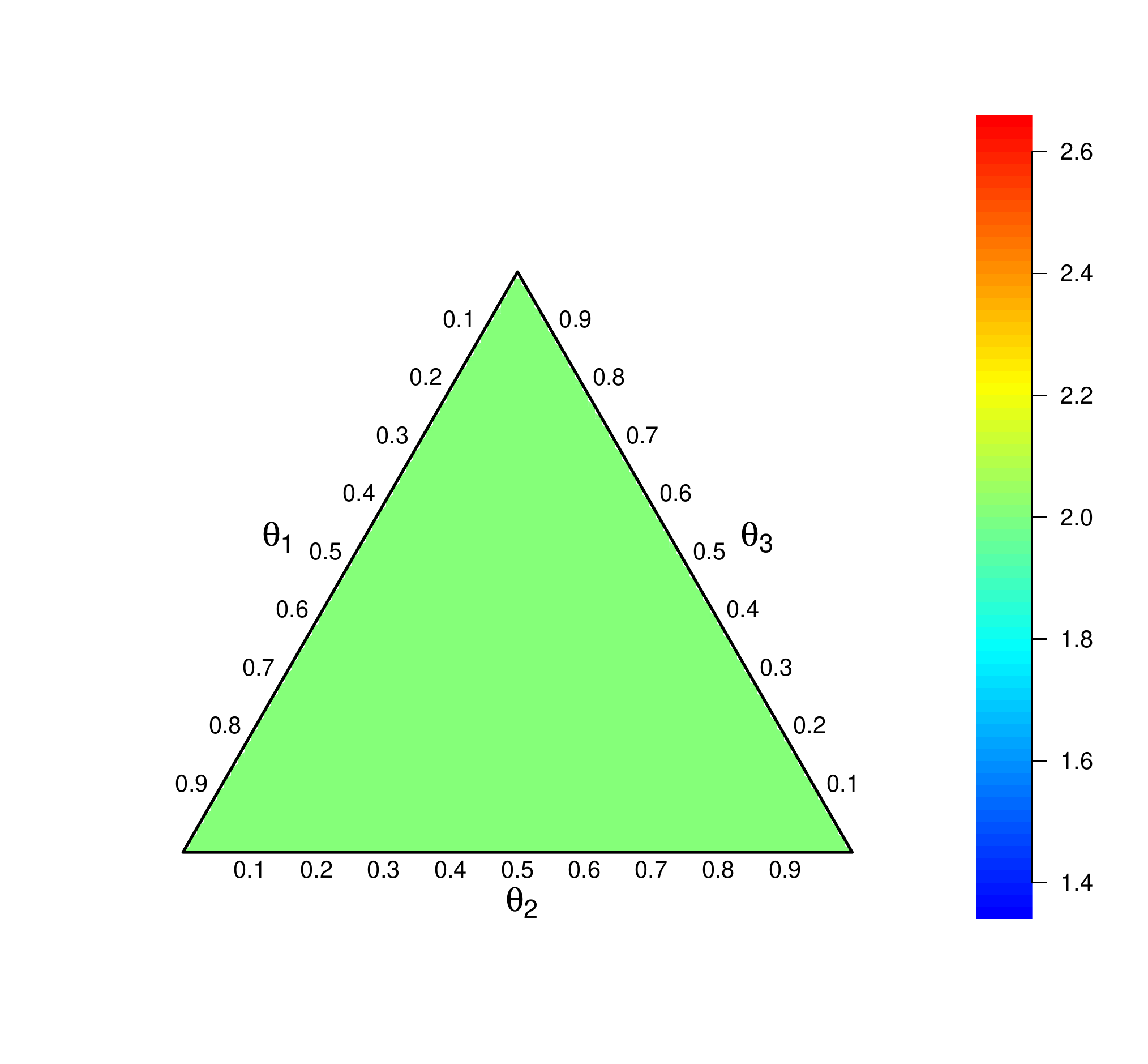}
 \includegraphics[scale=0.32]{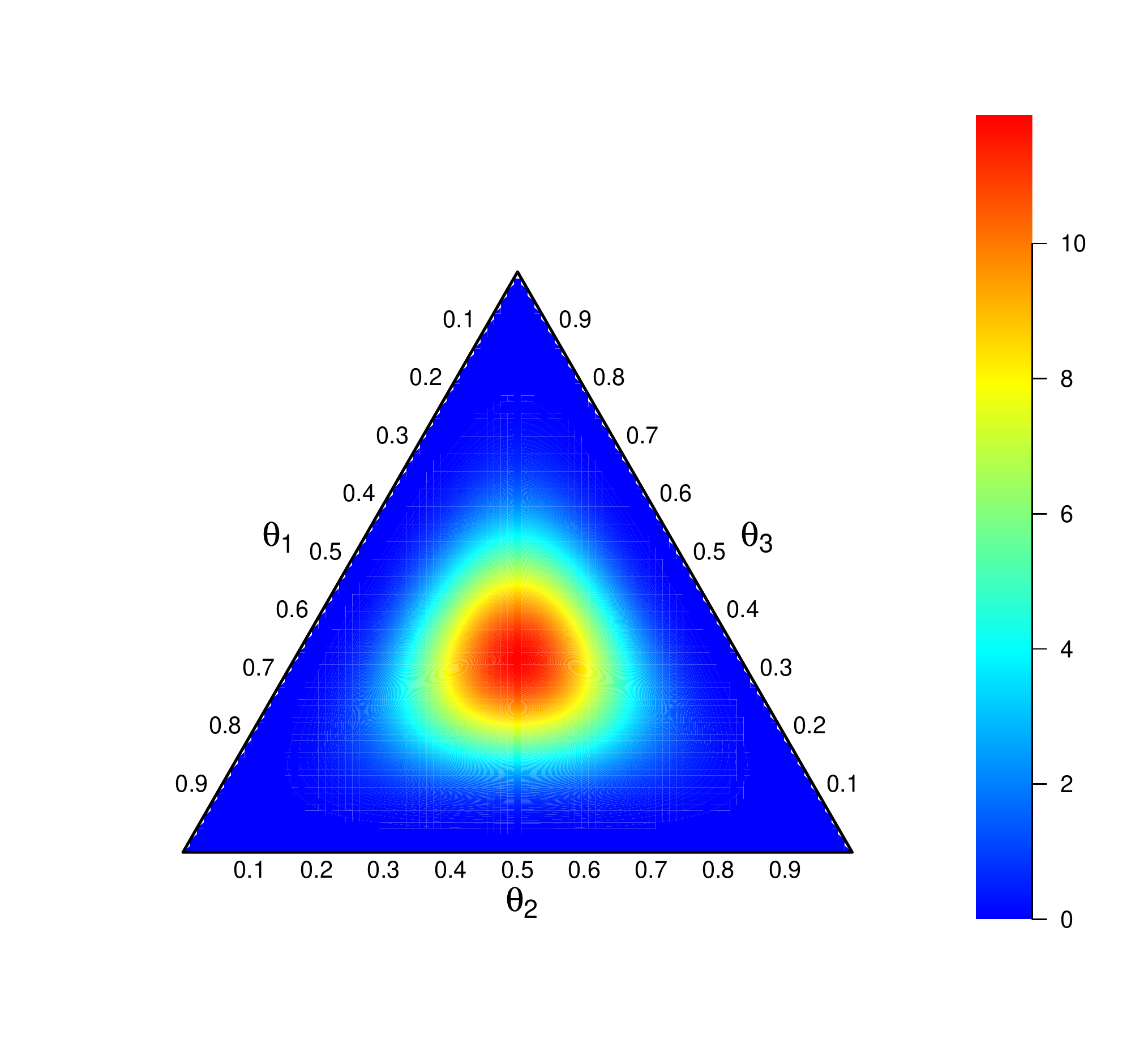}
 \includegraphics[scale=0.32]{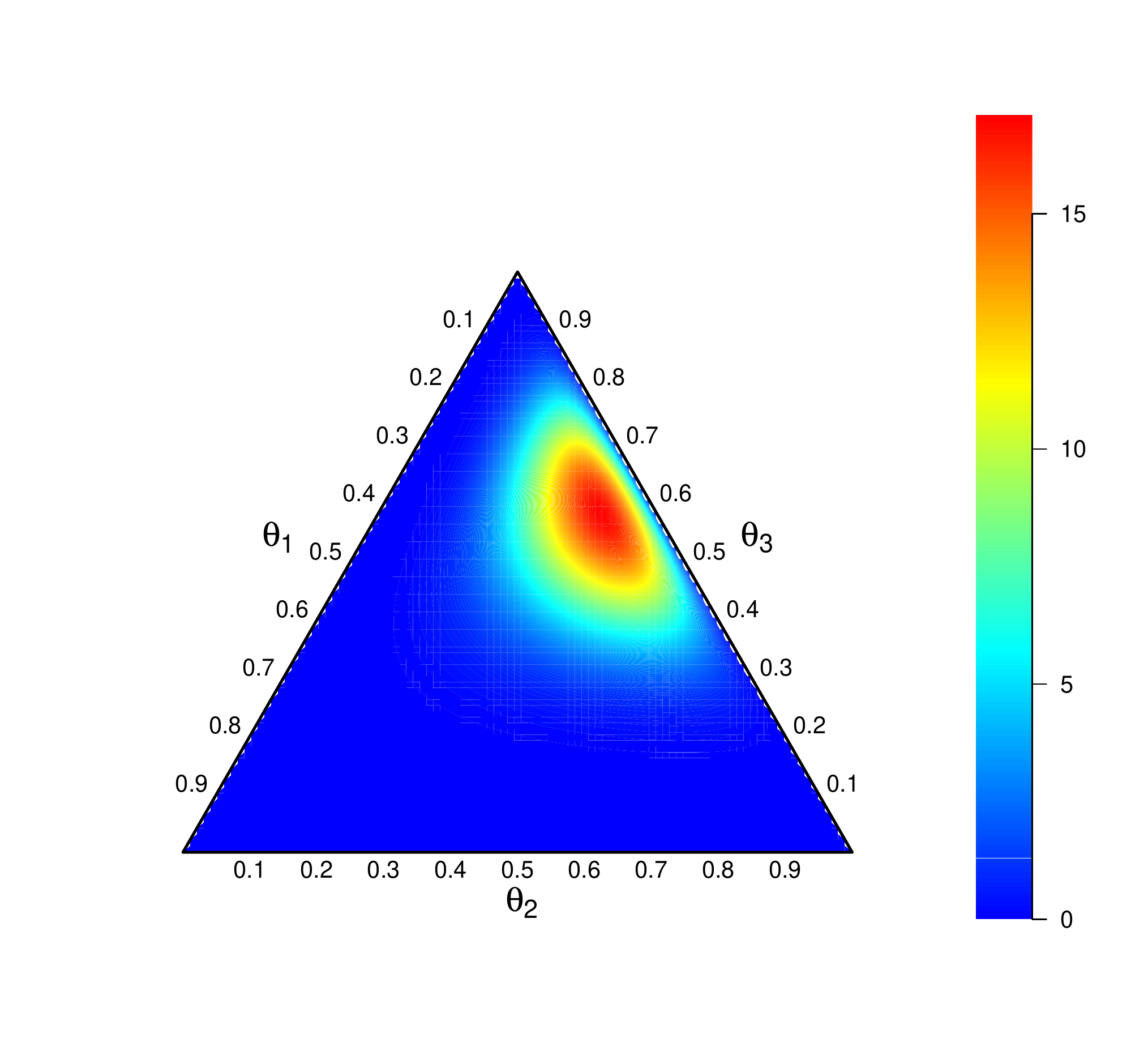}
 \includegraphics[scale=0.32]{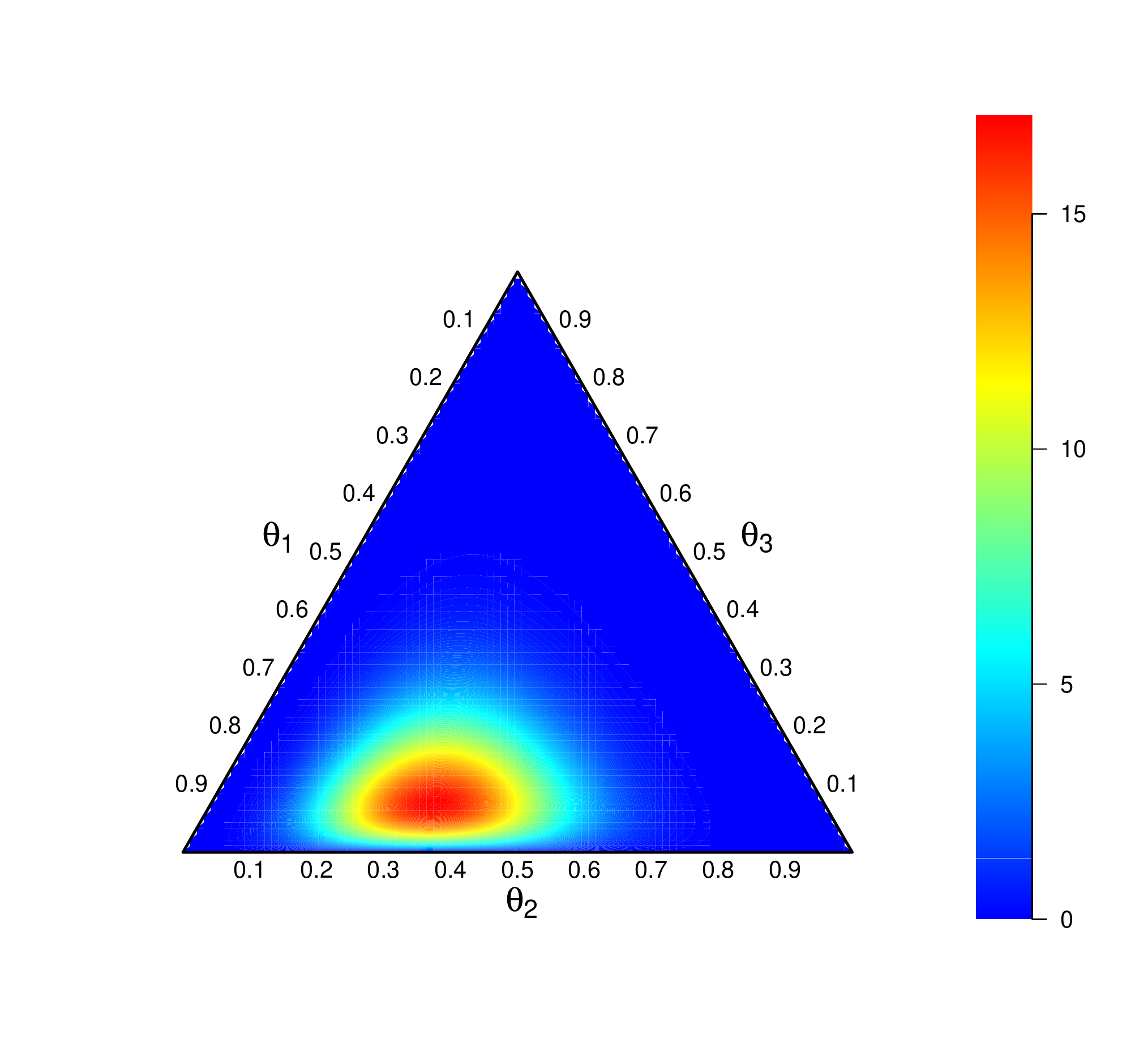}
 \caption{Ternary plots for the Dirichlet distribution with: $\bff{\alpha}= (1, 1, 1)$ and $\bff{\alpha}=(5, 5, 5)$ in the top row; $\bff{\alpha}=(2, 5, 8)$ and $\bff{\alpha}=(8, 5, 2)$ in the bottom row.}
 \label{fig-priors}
\end{figure}

We developed an R package \citep{R21} which provides a function to obtain point and interval estimates via Monte Carlo simulation as discussed in the previous section. Also, the package have a function to compute the minimum sample size to estimate the NPS through HPD interval via ALC (see Supplementary Material).

In Tables \ref{min-n-cen1}-\ref{min-n-cen4}, we present the minimum sample size to estimate the NPS using the HPD computed via ALC for all the scenarios for the prior distribution of $\bff{\theta}$ presented in Figure \ref{fig-priors} and some values of $\ell_{\text{max}}$ and $\rho$. For other scenarios the R package may be used. The cases where $\alpha_1=\alpha_2=\alpha_3=1$ and $\alpha_1=\alpha_2=\alpha_3=5$ represent scenarios in which the prior expected value of the NPS ($\Delta$) are equal to zero, but with different variability. The case where $\alpha_1=2$, $\alpha_2=5$ and $\alpha_3=8$ represents a scenario where the prior expected value of the NPS ($\Delta$) is positive, and where $\alpha_1=8$, $\alpha_2=5$ and $\alpha_3=2$ we have that the prior expected value of the NPS ($\Delta$) is negative, but the respective variances are equal.

For fixed $\rho$ ($\ell_{\text{max}}$), the minimum sample size decreases as $\ell_{\text{max}}$ ($\rho$) increases, as expected (Tables \ref{min-n-cen1}-\ref{min-n-cen4}). In the case where all the $\alpha_i$'s are equal and increase the minimum sample size seems to increases, irrespective the values of $\ell_{\text{max}}$ and $\rho$ (Tables \ref{min-n-cen1}-\ref{min-n-cen2}). The minimum sample size for the case where $\alpha_1=2$, $\alpha_2=5$ and $\alpha_3=8$ are approximately equal to those with $\alpha_1=8$ $\alpha_2=5$, $\alpha_3=2$ and the same $\ell_{\text{max}}$ and $\rho$ (Tables \ref{min-n-cen3}-\ref{min-n-cen4}), it makes sense since these scenarios are ``complementary'' with respect to the expected value but with the same variance.

For the adopted model parameters, the running time to compute the minimum sample size varied from 47 seconds to 4.69 hours, depending on the setting. The smaller the values of $\ell_{\text{max}}$ and/or $\rho$ the greater the running time. The computer that has been used has the following characteristics: OS Linux Ubuntu 20.04, RAM 7.7 GB, processor AMD PRO A8-8600B.

\begin{table}
\begin{center}
\caption{ALC based minimum sample size to estimate the NPS through the HPD with $\alpha_1=\alpha_2=\alpha_3=1$ for the prior distribution of $\bff{\theta}$.}\label{min-n-cen1}
\begin{tabular}{crrr}
\hline
 & \multicolumn{3}{c}{$\rho$} \\\cline{2-4}
$\ell_{\text{max}}$ & 0.01 & 0.05 & 0.10\\\hline
0.02 & 28199 & 16034 & 11328\\
0.04 & 7153  & 4085  & 2805 \\
0.06 & 3222  & 1808  & 1272 \\
0.08 & 1786  & 992   & 721  \\
0.10 & 1158  & 655   & 460  \\
0.12 & 807   & 460   & 315  \\
0.14 & 591   & 334   & 234  \\
0.16 & 452   & 249   & 179  \\
0.18 & 359   & 203   & 142  \\
0.20 & 291   & 164   & 114  \\
\hline
\end{tabular}
\end{center}
\end{table}

\begin{table}
\begin{center}
\caption{ALC based minimum sample size to estimate the NPS through the HPD with $\alpha_1=\alpha_2=\alpha_3=5$ for the prior distribution of $\bff{\theta}$.}\label{min-n-cen2}
\begin{tabular}{crrr}
\hline
 & \multicolumn{3}{c}{$\rho$} \\\cline{2-4}
$\ell_{\text{max}}$ & 0.01 & 0.05 & 0.10\\\hline
0.02 & 37737 & 21516 & 15157\\
0.04 & 9521  & 5396  & 3810 \\
0.06 & 4243  & 2386  & 1681 \\
0.08 & 2391  & 1330  & 944  \\
0.10 & 1515  & 860   & 599  \\
0.12 & 1046  & 590   & 409  \\
0.14 & 764   & 433   & 298  \\
0.16 & 585   & 326   & 224  \\
0.18 & 464   & 254   & 175  \\
0.20 & 371   & 204   & 139  \\
\hline
\end{tabular}
\end{center}
\end{table}

\begin{table}
\begin{center}
\caption{ALC based minimum sample size to estimate the NPS through the HPD with $\alpha_1=2$, $\alpha_2=5$ and $\alpha_3=8$ for the prior distribution of $\bff{\theta}$.}\label{min-n-cen3}
\begin{tabular}{crrr}
\hline
 & \multicolumn{3}{c}{$\rho$} \\\cline{2-4}
$\ell_{\text{max}}$ & 0.01 & 0.05 & 0.10\\\hline
0.02 & 28494 & 16105 & 11346\\
0.04 & 7116  & 4018  & 2816 \\
0.06 & 3198  & 1790  & 1258 \\
0.08 & 1779  & 1004  & 698  \\
0.10 & 1137  & 636   & 442  \\
0.12 & 788   & 442   & 307  \\
0.14 & 577   & 322   & 225  \\
0.16 & 433   & 243   & 166  \\
0.18 & 342   & 189   & 130  \\
0.20 & 275   & 150   & 100  \\
\hline
\end{tabular}
\end{center}
\end{table}

\begin{table}
\begin{center}
\caption{ALC based minimum sample size to estimate the NPS through the HPD with $\alpha_1=8$, $\alpha_2=5$ and $\alpha_3=2$ for the prior distribution of $\bff{\theta}$.}\label{min-n-cen4}
\begin{tabular}{crrr}
\hline
 & \multicolumn{3}{c}{$\rho$} \\\cline{2-4}
$\ell_{\text{max}}$ & 0.01 & 0.05 & 0.10\\\hline
0.02 & 28360 & 16040 & 11380\\
0.04 & 7160  & 4028  & 2861 \\
0.06 & 3190  & 1794  & 1263 \\
0.08 & 1795  & 1003  & 702  \\
0.10 & 1151  & 640   & 443  \\
0.12 & 781   & 440   & 308  \\
0.14 & 579   & 322   & 219  \\
0.16 & 438   & 240   & 164  \\
0.18 & 345   & 186   & 129  \\
0.20 & 278   & 152   & 100\\
\hline
\end{tabular}
\end{center}
\end{table}



%
%
%
%
%

\section{Illustrative example}\label{sec-example}

In the situation where the sample is not obtained yet, we may determine the sample size to obtain a HPD interval for the NPS via ALC by setting $\alpha_1$, $\alpha_2$, $\alpha_3$, $\ell_{\text{max}}$ and $\rho$. For example, if we consider $\alpha_1=\alpha_2=\alpha_3=1$, $\ell_{\text{max}}=0.10$ and $\rho=0.05$, the minimum sample size is 655 (Table \ref{min-n-cen1}), {\it i.e.}, to obtain a HPD interval with maximum length equals to 0.10 and respective probability equals to 0.95, we should ask 655 people the NPS question and then categorize them into detractors, passives and promoters in order to obtain the observed values of $X_1$, $X_2$ and $X_3$, respectively.

Given the difficult to obtain a real NPS dataset from a company because such a information is very sensitive, we consider a hypothetical dataset on financial services in three markets in the year of 2021 (see Supplementary Material) to mimic the application of the methods in a real situation.

To illustrate the methodology and the Bayesian updating, we consider the data from the first and second quarter of the Mexico market. For the first quarter we have no prior knowledge, then we set $\alpha_1=\alpha_2=\alpha_3=1$. For this quarter the numbers of detractors, passives and promoters are 136, 82 and 188, respectively, which implies a posterior Dirichlet distribution with vector parameter $\bff{\alpha}^*=(137, 83, 189)^\top$ for $\bff{\theta}$. Drawing a sample from this posterior distribution and computing its summaries, we have that a point estimate for the NPS is 0.127 and the HPD 95\% interval is [0.038, 0.206]. For the second quarter, we may use the posterior parameter of the first quarter as the prior parameter for the current quarter, {\it i.e.}, $\alpha_1=137$, $\alpha_2=83$ and $\alpha_3=189$. For the second quarter the numbers of detractors, passives and promoters are 136, 82 and 188, respectively. In this case, a point estimate for the NPS is 0.131 and the HPD 95\% interval is [0.072, 0.192]. All these results were obtained via the R package. Another simple way to obtain point and interval estimates for the NPS is to obtain (\ref{post-mean}) and (\ref{post-var}) for this data, as discussed in Section \ref{sec-bayes-model}.

\section{Concluding remarks}\label{sec-remarks}

For the first time in the literature the sample size determination for estimating the NPS is discussed. To approach this problem we consider a Bayesian approach via a multinomial/Dirichlet model and the average length criterion. We provide point and interval estimators for the NPS as in closed forms or via drawing a sample from the posterior distribution of the NPS and computing its summaries. Also, the Bayesian approach makes the inference updating becomes straightforward as illustrated in Section \ref{sec-example}, {\it i.e.}, a sequential procedure to estimate the NPS. Computational tools were developed to use these methodologies in practice.


\section*{Supplementary Material}

The Excel spreadsheet is available at \url{https://doi.org/10.5281/zenodo.7679211}.The R package is available at \url{https://github.com/eliardocosta/BayesNPS} (DOI: 10.5281/zenodo.7617770). The data used in the illustrative example is available at \url{https://www.kaggle.com/code/charlottetu/net-promoter-score/}.




\bibliographystyle{biometrika}
\bibliography{bibliografy}

\begin{thebibliography}{11}
\expandafter\ifx\csname natexlab\endcsname\relax\def\natexlab#1{#1}\fi

\bibitem[{Chen \& Shao(1999)}]{ChenShao99}
\textsc{Chen, M.-H.} \& \textsc{Shao, Q.-M.} (1999).
\newblock Monte {C}arlo estimation of {B}ayesian credible and {HPD} intervals.
\newblock \textit{Journal of Computational and Graphical Statistics}
  \textbf{8}, 69--92.

\bibitem[{Costa et~al.(2021)Costa, Paulino \& Singer}]{Costaetal21}
\textsc{Costa, E.~G.}, \textsc{Paulino, C.~D.} \& \textsc{Singer, J.~M.}
  (2021).
\newblock Sample size for estimating organism concentration in ballast water:
  {A Bayesian} approach.
\newblock \textit{Brazilian Journal of Probability and Statistics} \textbf{35},
  158 -- 171.

\bibitem[{Eskildsen \& Kristensen(2011)}]{EskildsenKristensen11}
\textsc{Eskildsen, J.~K.} \& \textsc{Kristensen, K.} (2011).
\newblock The accuracy of the {Net Promoter Score} under different
  distributional assumptions.
\newblock In \textit{2011 International Conference on Quality, Reliability,
  Risk, Maintenance, and Safety Engineering}.

\bibitem[{Keiningham et~al.(2008)Keiningham, Aksoy, Cooil, Andreassen \&
  Williams}]{Keininghametal2008}
\textsc{Keiningham, T.~L.}, \textsc{Aksoy, L.}, \textsc{Cooil, B.},
  \textsc{Andreassen, T.~W.} \& \textsc{Williams, L.} (2008).
\newblock A holistic examination of {Net Promoter}.
\newblock \textit{Journal of Database Marketing \& Customer Strategy
  Management} \textbf{15}, 79--90.

\bibitem[{Kristensen \& Eskildsen(2014)}]{KristensenEskildsen14}
\textsc{Kristensen, K.} \& \textsc{Eskildsen, J.} (2014).
\newblock Is the {NPS} a trustworthy performance measure?
\newblock \textit{The TQM Journal} \textbf{26}, 202--214.

\bibitem[{Markoulidakis et~al.(2021)Markoulidakis, Rallis, Georgoulas,
  Kopsiaftis, Doulamis \& Doulamis}]{Markoulidakisetal2021}
\textsc{Markoulidakis, I.}, \textsc{Rallis, I.}, \textsc{Georgoulas, I.},
  \textsc{Kopsiaftis, G.}, \textsc{Doulamis, A.} \& \textsc{Doulamis, N.}
  (2021).
\newblock Multiclass confusion matrix reduction method and its application on
  {Net Promoter Score} classification problem.
\newblock \textit{Technologies} \textbf{9}, 81.

\bibitem[{{R Core Team}(2022)}]{R21}
\textsc{{R Core Team}} (2022).
\newblock \textit{R: A Language and Environment for Statistical Computing}.
\newblock R Foundation for Statistical Computing, Vienna, Austria.

\bibitem[{Reichheld(2003)}]{Reichheld2003}
\textsc{Reichheld, F.~F.} (2003).
\newblock The one number you need to grow.
\newblock \textit{Harvard Business Review} \textbf{81}, 46--55.

\bibitem[{Rocks(2016)}]{Rocks2016}
\textsc{Rocks, B.} (2016).
\newblock Interval estimation for the ``{Net Promoter Score}".
\newblock \textit{The American Statistician} \textbf{70}, 365--372.

\bibitem[{Rossi \& Allenby(2003)}]{RossiAllenby03}
\textsc{Rossi, P.~E.} \& \textsc{Allenby, G.~M.} (2003).
\newblock Bayesian statistics and marketing.
\newblock \textit{Marketing Science} \textbf{22}, 304--328.

\bibitem[{Turkman et~al.(2019)Turkman, Paulino \& M{\"u}ller}]{Turkmanetal2019}
\textsc{Turkman, M. A.~A.}, \textsc{Paulino, C.~D.} \& \textsc{M{\"u}ller, P.}
  (2019).
\newblock \textit{Computational Bayesian statistics: an introduction}.
\newblock Cambridge: Cambridge University Press.

\end{thebibliography}

\end{document}